\documentstyle[12pt]{article}

\begin{document}

\rightline{DFPD 94/TH/31, May 1994}

\vspace{0.6cm}

{\begin{center}
{\large \bf A CLASS OF QUANTUM STATES WITH \\

\vspace{0.2cm}

 CLASSICAL-LIKE EVOLUTION} \\

\vspace{1.3cm}

{\large Salvatore De Martino\footnote{Electronic Mail:
demartino@vaxsa.dia.unisa.it} and Silvio De Siena\footnote{Electronic
Mail: desiena@vaxsa.dia.unisa.it}} \\

\vspace{0.2cm}

{\it Dipartimento di Fisica, Universit\`{a} di Salerno, \\
     and INFN, Sezione di Napoli, Gruppo collegato di Salerno, \\
     84081 Baronissi, Italia}

\vspace{0.7cm}

{\large Fabrizio Illuminati\footnote{Electronic Mail:
illuminati@mvxpd5.pd.infn.it}} \\

\vspace{0.2cm}

{\it Dipartimento di Fisica ``G. Galilei", Universit\`{a} di Padova, \\
     and INFN, Sezione di Padova,  \\
     35131 Padova, Italia}

\end{center}}

\vspace{1cm}

{\begin{center} \large \bf Abstract \end{center}}

In the framework of the Stochastic Formulation of Quantum Mechanics
we derive
non-stationary states for a class of time-dependent potentials.
The wave-packets follow a classical motion with
constant dispersion. The new states
define a possible extension of the
harmonic-oscillator coherent states.
As an explicit application we study  a sestic
oscillator potential.

\newpage

Coherent states were introduced for the first time by Schr\"odinger
in his attempt at finding the quantum mechanical states that
evolve according to a classical law without spreading of the
wave-packet [1].

Starting from the seminal works of Glauber,
Klauder and Sudarshan [2], coherent states have come to
play a central role in many different areas of research, ranging
from quantum mechanics and quantum optics to quantum field theory
and condensed matter physics. In the course of time they have been
thoroughly analysed in a group-theoretical framework [3], and
extended in the direction of squeezing phenomena [4].

On the other hand, the original problem formulated by Schr\"odinger,
namely whether and in what sense coherent structures can be introduced
for general quantum systems, is still open.

Coherent states of arbitrary potentials should be such to retain
as much as possible the basic property of the harmonic-oscillator
coherent states, namely  they should be non-stationary states
without dispersion of the wave-packet width
and with time evolution driven by a classical equation.
Thus the search for
a sensible definition of generalized coherent states should be
based on the above physical requirement. This way of looking at the
problem has been strongly advocated and pursued by Nieto and
collaborators who constructed coherent states associated
to arbitrary potentials ([5], [6]).

These coherent states, however, can only be constructed
for classically integrable systems for which there exists a set
of canonical coordinates and momenta $\{Q,P\}$ such that the respective
Hamiltonians can be reduced to quadratures. The canonical
transformations from the original variables $\{q,p\}$ to
the ``natural" [5] variables $\{Q,P\}$ are energy-dependent.

Quantization of the variables $\{Q,P\} \rightarrow \{ \hat{Q},
\hat{P} \}$ allows to built the coherent states.

The energy-dependence of the classical "natural" [5]
variables $\{Q,P\}$ determines the level-dependence
of $\hat{Q}$ and $\hat{P}$ through the
Hamiltonian $\hat{H}$ and prevents in general the exact solution of
the equations defining the generalized coherent states. Approximate
solutions can be obtained by letting $\hat{H} \rightarrow \langle
\hat{H} \rangle$. The more confining a potential is, and the closer
to equally spaced its energy eigenstates are, then the better the
generalized coherent state will evolve according to a classical
law, with conserved spreading of the wave-packet, until, in the limit
of systems with equally spaced energy levels, the approximation becomes
exact [6].

The above scheme has been applied also to define the
squeezed states of general potentials [7], while other
classes of generalized squeezed states have been recently introduced
by Nieto and Truax by a annihilation-operator method [8].
The different definitions of generalized coherent and squeezed states
for non-harmonic potentials and the connections among them are nicely
reviewed in a recent report by Nieto [9].

In this letter we consider again the problem of Schr\"odinger
for non-harmonic potentials. Our strategy to construct
generalized coherent states is based on imposing some
conditions that have to be satisfied directly in the
quantum regime and for the original variables $\{ \hat{q},
\hat{p} \}$. More precisely, the requirements we impose are
a constant spreading $\Delta \hat{q}$
and the following classical evolution equations
\[
m \frac{d^2}{dt^2} \langle \hat q \rangle
=-\nabla V(q,t) \, \; |_{q=\langle \hat q \rangle} \; ,
\]
\begin{equation}
\end{equation}
\[
\langle {\hat p} \rangle=m \frac{d}{dt} \langle {\hat q} \rangle \; .
\]
\noindent

Our construction
is obtained exploiting the techniques of Nelson stochastic
quantization and the relation among stochastic and quantum
uncertainties. We select a class of potentials
which satisfy Schr\"odinger's conditions; these potentials in general
depend from the classical trajectory described by the
configurational mean value.

The Stochastic Formulation of Quantum Mechanics (S.F.Q.M.),
originally introduced by Nelson,
is a useful tool in the study of the relationships
among classical and quantum evolutions [10].
It is indeed based on the
replacement of the deterministic trajectories of classical mechanics
with suitably chosen Brownian paths, and it is thus close in spirit
to Feynman path integral quantization.

Consider, without loss of
generality, a one-dimensional system. At the kinematical level S.F.Q.M.
associates to the classical configurational variable
a diffusion process $q(t)$ governed by Ito's stochastic
differential equation
\begin{equation}
dq(t) = v_{(+)}(q(t),t)dt + \left( \sqrt{\frac{\hbar}{2m}}
\right) dw(t) \, , \; \; \; \; dt > 0 \, \, .
\end{equation}

\noindent Here $v_{(+)}(q(t),t)$ is the forward drift, and
$dw(t)$ is the time-increment of the Gaussian white noise $w(t)$,
superimposed on the otherwise deterministic evolution, with
expectation $E(dw(t))=0$ and covariance $E(dw^{2}(t)) =2dt$.
The probability density $\rho(x,t)$ associated to the process satisfies
the forward and backward Fokker-Planck equations.
The forward and the backward drifts $v_{(+)}(x,t)$ and
$v_{(-)}(x,t)$ are defined as
\[
v_{(+)}(x,t) = \lim_{\Delta t \rightarrow 0^{+}}E
\left( \frac{q(t + \Delta t) - q(t)}{\Delta t}\left.
\right| q(t) = x \right) \, ,
\]
\begin{equation}
\end{equation}
\[
v_{(-)}(x,t) = \lim_{\Delta t \rightarrow 0^{+}}E
\left( \frac{q(t) - q(t - \Delta t)}{\Delta t}\left. \right|
q(t) = x \right) \, .
\]

{}From definitions (3) it is clear that
the forward (backward) drift is the mean
forward (backward) velocity field.

The relation between $v_{(+)}$ and $v_{(-)}$ is:
\begin{equation}
v_{(-)} = v_{(+)} -
\frac{\hbar}{m}\frac{\partial_{x}\rho}{\rho} \, \, .
\end{equation}

We define also the osmotic velocity $u(x,t)$ and
the current velocity $v(x,t)$ as
\[
u =  \frac{v_{(+)} - v_{(-)}}{2} = \frac{\hbar}{2m}
\frac{\partial_{x}\rho}{\rho} \, ,
\]
\begin{equation}
\end{equation}
\[
v=\frac{v_{(+)}+v_{(-)}}{2} \, .
\]

The osmotic velocity $u(x,t)$ then``measures" the
non-differentiability of the random trajectories; in the
 classical limit,
$\hbar  \rightarrow 0$, $u$
vanishes and the current velocity $v(x,t)$ goes to the
classical velocity.

As a last consequence of Fokker-Planck equation we have for the
probability density $\rho(x,t)$ the
continuity equation
\begin{equation}
\partial_{t}\rho = -\partial_{x}(\rho v) \, .
\end{equation}

At the dynamical level each single-particle quantum state $\Psi(x,t)$
written in the form
\begin{equation}
\Psi(x,t) = \rho^{\frac{1}{2}}\exp \left[ \frac{i}{\hbar}S(x,t) \right] \; ,
\end{equation}

\noindent where $\rho$ and $S$ are two real functions, corresponds in
S.F.Q.M. to the diffusion process $q(t)$ with
\begin{equation}
\rho(x,t) = |\Psi(x,t)|^{2} \; ,
\end{equation}

\noindent and
\begin{equation}
v(x,t) = \frac{1}{m}\partial_{x}S(x,t) \; ,
\end{equation}

\noindent where $m$ is the mass of the particle.

The Schr\"odinger equation with potential $V(x,t)$ for the complex
wave-function $\Psi$ is then equivalent to two coupled real
equations for the probability density $\rho$ and the phase $S$.

The first equation is the continuity equation (6), while the second
one is the hydrodynamical Hamilton-Jacobi-Madelung (HJM) equation
\begin{equation}
\partial_{t}S + \frac{(\partial_{x}S)^{2}}{2m} - \frac{\hbar^{2}}
{2m}\frac{\partial_{x}^{2}\sqrt{\rho}}{\sqrt{\rho}} = - V \; ,
\end{equation}

\noindent or in terms of the osmotic velocity $u$
\begin{equation}
\partial_{t}S + \frac{m}{2}v^{2} - \frac{m}{2}u^{2} - \frac{\hbar}{2}
\partial_{x}u = - V \; .
\end{equation}

The gradient of the HJM equation yields
\begin{equation}
m\partial_{t}v + mv\partial_{x}v -mu\partial_{x}u -\frac{\hbar}{2}
\partial_{x}^{2}u = - \partial_{x}V \; .
\end{equation}

This equation can be seen as the quantum analogue of Newton's
equation in a force field [10].

The correspondence between expectations and correlations defined
in the stochastic and in the canonic pictures are

\newpage

\[
\langle \hat{q} \rangle = E(q) \, , \; \; \; \; \langle \hat{p}
\rangle = mE(v) \, ,
\]

\begin{equation}
\Delta \hat{q} = \Delta q \, ,
\end{equation}

\[
(\Delta \hat{p})^{2} = m^{2}[(\Delta u)^{2} + (\Delta v)^{2}] \, ,
\]

\noindent and furthermore  the chain inequality  holds
\begin{equation}
(\Delta \hat{q})^{2} (\Delta \hat{p})^{2} \geq
m(\Delta q)^{2} (\Delta u)^{2} \geq \frac{{\hbar}^{2}}{4} \, ,
\end{equation}

\noindent where $\hat{q}$ and $\hat{p}$ are the position and
momentum observables in the Schr\"odinger picture, $\langle \cdot
\rangle$ denotes the expectation value of the operators in the
given state $\Psi$, $E(\cdot)$ is the expectation values of the
stochastic variable associated in the Nelson picture to the
state $\{\rho, S\}$, and $\Delta(\cdot)$ denotes the root mean
square deviation.

The chain inequality (14), i.e. the osmotic
velocity-position stochastic
uncertainty and its equivalence with the momentum-position
quantum uncertainty, were proven in [11].

The diffusion processes that minimize the
stochastic uncertainty product
have been recently derived [12].
{}From  eqs. (13), (14) and the analysis exploited in ref. [12],
it follows that the Nelson
minimum uncertainty states are
comprehensive both of the standard Glauber
and Klauder-Sudarshan harmonic oscillator coherent states
(Heisenberg minimum uncertainty states), and of all the
Schr\"odinger minimum uncertainty states.

Through saturation of the osmotic velocity-position stochastic
uncertainty we have
\begin{equation}
u(x,t)=-\frac{\hbar}{2m\Delta q}\left( \frac{x-E(q)}{\Delta q}\right) \; ;
\end{equation}

\noindent if we insert eq. (15) in the continuity equation (6) we
obtain, after
simple calculations, the following form of the current velocity

\begin{equation}
v(x,t)=\frac{d}{dt}E(q) + \left( \frac{x-E(q)}{\Delta q} \right)
\frac{d}{dt}\Delta q \; .
\end{equation}

The minimum uncertainty stochastic states (MUSSs) are thus determined
by a current velocity $v$ and an osmotic velocity $u$ both linear
in the argument $(x-E(q))/\Delta q$. It can be shown that
the MUSSs can be
divided in two classes, the first with constant $\Delta q$ and
the second with time-dependent $\Delta q$.
In the actual context
it is convenient to consider $\Delta q$ constant (for the more
general case see [12]).
Inserting, relations (15) and (16)
into eq. (11) we find that
\begin{equation}
V(x,t) = \frac{m}{2}\omega^{2}x^{2} ,
\end{equation}

\begin{equation}
\frac{d^2}{dt^2}E(q) = - \omega^{2}E(q) ,
\end{equation}
\noindent
and we obtain the standard Glauber coherent states wave functions.
We note that, while $u(x,t)$ of the form (15) implies $v(x,t)$ of
the form (16), the contrary is not true.
In fact, this is an interesting general feature of the S.F.Q.M.:
a given choice of the current velocity
$v$ determines a whole class of osmotic velocities $u$
and thus selects a class of quantum states with
the same dynamical evolution. A simple but relevant example
[13]
is given by the choice $v=0$; in this case it is immediately seen that
eq. (11) becomes the standard eigenvalue equation for the
Hamiltonian operator.

Now, we use this property to construct general coherent states in the
sense of Schr\"odinger.
To this aim, it is natural to search for the class of quantum
states selected by a current velocity of the minimum-uncertainty form
(16) with constant $\Delta q$:
\begin{equation}
v=\frac{d}{dt}E(q) \, .
\end{equation}
Exploiting the
continuity equation (6) written in terms of $v$ and $u$
and substituting for $v$ the form (16)
we have that $u$ satisfies the equation
\begin{equation}
\partial_{t}u = -E(v)\partial_{x}u \, ,
\end{equation}

\noindent whose general solution  is

\begin{equation}
u=\frac{1}{\Delta q}G(\xi) \; ,
\end{equation}

\noindent with
\begin{equation}
\xi = \frac{x-E(q)}{\Delta q} \; .
\end{equation}

The function $G(\xi)$ can be arbitrary, only restricted by
the condition that $u$ must yield
a normalizable probability density
$\rho(\xi)$.

The minimum-uncertainty osmotic velocity (15) linear
in $\xi$ is then just a particular case of the
general form (22).

Inserting (21) and (22) into eq. (12) we obtain the identity
\begin{equation}
m\frac{d^{2}}{dt^{2}}E(q) = \frac{m}{{(\Delta q)^{3}}}\left(
\frac{\hbar}{2m}G''(\xi) + G(\xi)
G'(\xi) \right) - \partial_{x}V(x,E(q)) \; ,
\end{equation}
\noindent where the primes denote derivation respect to $\xi$.
If one compute relation (23)  in
$x =E(q) \equiv \langle {\hat q} \rangle$
(which corresponds also to $\xi=0$),
it is then evident that the classical equations
of motion (1) for E(q)
and then for $\langle \hat{q} \rangle$
is guaranteed by the choice (19) and the condition
\begin{equation}
\frac{m}{{(\Delta q)^{3}}} \left. \left(
\frac{\hbar}{2m}G''(\xi) + G(\xi)
G'(\xi) \right) \right|_{\xi=0} = 0 .
\end{equation}

Now, a particular choice of  $G(\xi)$ which satisfies condition (24)
completely determines, through the relations and definitions
(8), (9), (11), (12) the potential $V$
and the wave-function of the "coherent-like"
state, which has namely the general form

\begin{equation}
\Psi_{c}(x,t)=\frac{1}{\Delta \hat{q}}\exp\left\{ \frac{2m}{\hbar
\Delta \hat{q} }\int_{0}^{\xi}G(\xi')d\xi' + \frac{i}{\hbar}\left(
x\langle \hat{p} \rangle + S_{0}(t) \right) \right\} \, ,
\end{equation}

\noindent where $S_{0}(t)$ is  calculated in
$x=x_{0}=0$.

To clarify our structure, we now choose a particular form
of the function $G(\xi)$, and namely of the osmotic velocity $u$:
\begin{equation}
u=-\frac{\hbar}{2m\Delta q}\xi^{3} \, .
\end{equation}

The choice of the coefficient $\hbar/2m$ is dictated by dimensional
considerations,
while the minus sign is fixed by requiring normalization of the density
$\rho$. Notice that it is the same coefficient appearing in the
Heisenberg minimum-uncertainty $u$ . In fact,
$-\hbar/2m$ is the universal coefficient for any polynomial choice
of $u(\xi)$.
{}From eq. (27) and the first of eqs. (5) one obtains the normalized
probability density
\begin{equation}
\rho(x,t)=
\sqrt{\frac{2}{(\Delta q)^{2}}}\left[ \Gamma \left( \frac{1}{4}
\right) \right]^{-1}\exp \left[ -\frac{(x-E(q))^{4}}{4(\Delta q)^{4}}
\right] \, ,
\end{equation}

\noindent where  $\Gamma (1/4)$ is the Gamma-function
 $\Gamma (s)$ evaluated in $s=1/4$.
Inserting the choice (26) for $u$ and the expression (19) for $v$
in the HJM equation (11) we obtain that the potential must
be of the form:
\begin{equation}
V(x,t) = \sum_{i=0}^{6}a_{i}(E(q))x^{i} \; .
\end{equation}

This is the expression of a sestic oscillator potential, whose
coefficients $\{a_{i}\}$ are time-dependent
as functions of the expectation $E(q)$.
Their explicit expressions are computed through the identity
(11) with the choices (26), (19), and are given by:
\begin{eqnarray}
a_{6} & = & \frac{\hbar^{2}}{8m(\Delta q)^{8}} \; \; \; \;
a_{5} = -\frac{3\hbar^{2}}{4m(\Delta q)^{8}}E(q) \; , \nonumber \\
&& \nonumber \\
a_{4} & = & \frac{15\hbar^{2}}{8m(\Delta q)^{8}}E^{2}(q) \; \; \; \;
a_{3} = -\frac{5\hbar^{2}}{2m(\Delta q)^{8}}E^{3}(q) \; , \nonumber \\
&& \nonumber \\
a_{2} & = & \frac{15\hbar^{2}}{8m(\Delta q)^{8}}E^{4}(q) -
\frac{3\hbar^{2}}{4m(\Delta q)^{4}} \; , \\
&& \nonumber \\
a_{1} & = & -m\frac{d^{2}}{dt^{2}}E(q) + \frac{3\hbar^{2}}
{2m(\Delta q)^{4}}E(q) - \frac{3\hbar^{2}}{4m(\Delta q)^{8}}E^{5}(q) \; ,
\nonumber \\
&& \nonumber \\
a_{0} & = & - \frac{d}{dt}S_{0}(t) - \frac{m}{2}E^{2}(v)
+ \frac{\hbar^{2}}{8m(\Delta q)^{8}}E^{6}(q) - \frac{3\hbar^{2}}
{4m(\Delta q)^{4}}E^{2}(q) \; . \nonumber
\end{eqnarray}

The coefficients are thus all time-dependent but that of the highest
power, which is constant. The coefficient of the quadratic power is
insteed made up of two contributions, a constant and a time-dependent
one. It is straightforward to verify that the  condition
 (24) is identically satisfied. We are thus assured that
the states corresponding to the osmotic velocity (26) follow
the classical evolution (1).
{}From the coefficient of the linear term in the potential, we
read off the actual equation of motion obeyed by the
expectation  $\langle \hat{q} \rangle \equiv E(q)$ and so determine the
corresponding
time-evolution of the state.
By choosing without loss
of generality  $a_{0}=a_{1} = 0$, one can easily verify that
the time-dependent part of the potential (28) gives no contribution to
the classical equation of motion (1), which reads
\begin{equation}
m\frac{d^{2}}{dt^{2}}\langle \hat{q} \rangle
= -\frac{3\hbar^{2}}{4m(\Delta \hat{q})^{8}} {\langle \hat{q} \rangle}^{5}
+ \frac{3\hbar^{2}}{2m(\Delta \hat{q})^{4}} \langle \hat{q} \rangle \; .
\end{equation}

The state thus evolves in a classical potential field $V_{cl}$
which is obtained
by retaining only the time-independent part of
the expression $V$, and reads
\begin{equation}
V_{cl}(x) = \frac{\hbar^{2}}{8m(\Delta \hat{q})^{8}}x^{6} -
\frac{3\hbar^{2}}{4m(\Delta \hat{q})^{4}}x^{2} \; .
\end{equation}

The classical part $S_0(t)$ of the action $S(x,t)$ is readily determined
from the last of eqs. (29) by exploiting
eq. (30).
Collecting things together, reminding definitions (7), (13)
and expression (27)
for the density $\rho$, one finally obtains
the explicit expression for the states:
\begin{equation}
\Psi_{c}(x,t)= \left[ \frac{\Delta \hat{q}
\Gamma(1/4)}{2} \right]^{-\frac{1}{4}}\exp\left\{ -\frac{(x -
\langle \hat{q} \rangle )^{4}}{8(\Delta \hat{q})^{4}}
+ \frac{i}{\hbar}\left( \langle \hat{p} \rangle x -\frac{1}{2}
\langle \hat{q} \rangle \langle \hat{p} \rangle \right) \right\} \; .
\end{equation}

The wave-functions (32) appear as a natural extension of the standard
Glauber coherent states to non-harmonic potentials. The modulus is
modulated
by the actual interaction. The phase retains the same structure
as in the case of
the Glauber wave-functions; namely, all these phases in fact encode
the classical-like dynamics of the wave packet.

Note that we give here the sestic polynomial as the  simplest solvable
example besides the well known harmonic one. However, there
are many other potentials with classical motion that can be
selected through our approach; for instance, more complex
polynomials, periodic potentials, an extension of the Morse potential, etc.

In conclusion, we have found a class of potentials and
states compatible with condition (1) and constant $\Delta {\hat q}$.

Our class of states contains as a subset the standard Glauber
coherent states.

It is interesting to observe, as it can be
seen from the above polynomial example (26)-(32), that these
states are associated to potentials which are time-dependent through
$\langle {\hat q}\rangle$. This last quantity is in general
governed by condition
(1) which, in the case of the sestic oscillator potential,
specializes to eq. (30).

We think that these new states show some intriguing features
and should deserve further study and understanding.
A possible future line of research might be connected with
the fact that
coherent states are a natural bridge for studying the quantum-
classical corrispondence. From this point of view, the
construction of coherent-like states for non-harmonic
potentials might be useful to understand the
manifestation of classical chaos in quantum systems [14].

\end{document}